\documentclass[fleqn,twoside]{article}
\usepackage{espcrc2}

\usepackage{graphicx}
\usepackage[figuresright]{rotating}
\usepackage{graphics}
\usepackage{cite}   

\newcommand{\lt}{\left}
\newcommand{\rt}{\right}
\newcommand{\no}{\nonumber}
\newcommand{\nn}{\nonumber \\}
\newcommand{\ov}[1]{\overline{#1}}
\newcommand{\eq}[1]{Eq.~(\ref{#1})}

\newcommand{\imag}{\mathrm{Im}\,}
\newcommand{\real}{\mathrm{Re}\,}

\newcommand{\Bbar}{\,\overline{\!B}}

\newcommand{\bbd}{\ensuremath{B_d\!-\!\Bbar{}_d\,}}
\newcommand{\bbs}{\ensuremath{B_s\!-\!\Bbar{}_s\,}}
\newcommand{\bbq}{\ensuremath{B_q\!-\!\Bbar{}_q\,}}
\newcommand{\bb}{\ensuremath{B\!-\!\Bbar\,}}
\newcommand{\bbm}{\bb\ mixing}

\newcommand{\bbms}{\bbs\ mixing}
\newcommand{\bbmd}{\bbd\ mixing}
\newcommand{\BsorBsbar}{\raisebox{7.7pt}{$\scriptscriptstyle(\hspace*{8.5pt})$}
  \hspace*{-10.7pt}\!\Bbar_{s}} 
\newcommand{\BqorBqbar}{\raisebox{7.7pt}{$\scriptscriptstyle(\hspace*{8.5pt})$}
  \hspace*{-10.7pt}\!\Bbar_{q}} 
 
\newcommand{\bra}[1]{\ensuremath{\langle #1 |}}
\newcommand{\ket}[1]{\ensuremath{| #1 \rangle }}
\newcommand{\fig}[1]{Fig.~\ref{#1}}

\newcommand{\lbar}{\ov{\Lambda}}
\newcommand{\dm}{\ensuremath{\Delta M}}
\newcommand{\dg}{\ensuremath{\Delta \Gamma}}
\newcommand{\epm}[2]{
 \raisebox{-0.5ex}{\shortstack[l]{$\scriptstyle+#1$\\$\scriptstyle-#2$}}}

\title{
\vspace{-3cm}{\small
TTP-06-33\hfill hep-ph/0612310}\\[2.5cm]
$B_d$ and $B_s$ mixing: mass and width differences and CP
  violation\thanks{in collaboration with Alexander Lenz.
Talk at \emph{Beauty 2006}, 25-29 Sep 2006, Oxford, England.}}

\author{Ulrich Nierste\\         
        Institut f\"ur Theoretische Teilchenphysik (TTP)\\
        Karlsruhe Institute of Technology --- Universit\"at Karlsruhe\\
        76128 Karlsruhe\\
        Germany}
       
\begin{document}

\begin{abstract}
  \bbm\ involves three physical parameters: the magnitudes of the
  off-diagonal elements of the mass and decay matrices and their
  relative phase.  They are related to the mass and width differences
  between the mass eigenstates and to the CP asymmetry in
  flavour-specific decays, $a_{\rm fs}$. Introducing a new operator
  basis I present new, more precise theory predictions for the width
  differences in the $B_s$ and $B_d$ systems: in the Standard Model one
  finds $\dg_s = 0.088 \pm 0.017\, \mbox{ps}^{-1}$ and $\dg_d = ( 26.7
  \epm {5.8}{6.5} ) \cdot 10^{-4} \; {\rm ps}^{-1}$.  Updates of the
  mass differences $\dm_d$ and $\dm_s$ and of $a_{\rm fs}^d$ and $a_{\rm
    fs}^s$ are also presented.  Then I discuss how various present and
  future measurements can be combined to constrain new physics. The
  extraction of a new CP phase $\phi_s^\Delta$ from data on $a_{\rm
    fs}^s$ also profits from our new operator basis.  Confronting our
  new formulae with D\O\ data we find that $\sin \phi_s^\Delta$ deviates
  from zero by $2\sigma$.  \vspace{1pc}
\end{abstract}

\maketitle

\section{\bbm\ basics}
Loop-induced $b\to s$ transitions are currently receiving much attention
from experiment and theory. Present data leave room for new physics with
sources of flavour beyond the Cabibbo-Kobayashi-Maskawa (CKM)
mechanism \cite{ckm} of the Standard Model (SM).  For example, an extra
contribution to $b\to s \ov{q} q$, $q=u,d,s$, decay amplitudes with a
new CP phase can alleviate the $\sim 2.6\sigma$ discrepancy between the
measured mixing-induced CP asymmetries in these $b\to s$ penguin modes
and the Standard Model prediction \cite{ichep06}. If this new
contribution involves $b$ and $s $ quarks with the same chirality, no
tension with the well-measured $b \to s \gamma$ branching fraction
occurs.  Supersymmetric grand-unified models can naturally accommodate
such new $b \to s$ amplitudes \cite{cmm}: right-handed quarks reside in
the same quintuplets of SU(5) as left-handed neutrinos, so that the
large atmospheric neutrino mixing angle could well affect squark-gluino
mediated $b\to s$ transitions \cite{jn}.  Obviously \bbms\ plays an
important role in the search for new physics in $b\to s$ FCNC's. In this
talk I present an update of the theory predictions of the physics
observables related to \bbms. Since the formalism equally applies to 
\bbmd, the corresponding quantities for the $B_d$ system are also
presented. The presented results are derived with the use of a novel
operator basis, which leads to a better precision of the theory
predictions. Details can be found in \cite{ln}.    

\bbq\ oscillations are governed by a Schr\"odinger
equation
\begin{equation}
i \frac{d}{dt}
\left(\!
\begin{array}{c}
\ket{B_q(t)} \\ \ket{\ov{B}_q (t)}
\end{array}
\! \right)
=
\left( M^q - \frac{i}{2} \Gamma^q \right)
\left(\!
\begin{array}{c}
\ket{B_q(t)} \\ \ket{\ov{B}_q (t)} 
\end{array}
\!\right)\label{sch}
\end{equation} 
with the mass matrix $M^q$ and the decay matrix $\Gamma^q$ and $q=d$ or
$q=s$.  The physical eigenstates $\ket{B_H}$ and $\ket{B_L}$ with the
masses $M^q_H,\,M^q_L$ and the decay rates $\Gamma^q_H,\,\Gamma^q_L$ are
obtained by diagonalizing $M^q-i \Gamma^q/2$.  The \bbq\ oscillations in
\eq{sch} involve the three physical quantities $|M_{12}^q|$,
$|\Gamma_{12}^q|$ and the CP phase
$\phi_q=\arg(-M_{12}^q/\Gamma_{12}^q)$ (see e.g.\ \cite{run2}).  The
mass and width differences between $B_{L}$ and $B_{H}$ are related to
these quantities as
\begin{eqnarray}
\dm_q &=& M^q_H -M^q_L \; = \; 2\, |M_{12}^q|, \label{dmdg1} \\
\dg_q & = & \Gamma^q_L-\Gamma^q_H \; =\;
        2\, |\Gamma_{12}^q| \cos \phi_q, \label{dmdg2}
\end{eqnarray}
up to numerically irrelevant corrections of order $m_b^2/M_W^2$.  
$\dm_q$ simply equals the frequency of the \bbq\ oscillations.
A third quantity providing independent information on the 
mixing problem in \eq{sch} is
\begin{eqnarray}
a^q_{\rm fs}
     &=&
    \imag \frac{\Gamma_{12}^q}{M_{12}^q}
    \; = \; \frac{|\Gamma_{12}^q|}{|M_{12}^q|} \sin \phi_q
 . \label{defafs}
\end{eqnarray}
$a_{\rm fs}^q$ is the CP asymmetry in \emph{flavour-specific} $B_q\to f$
decays and quantifies \emph{CP violation in mixing}. The standard way to
measure $a_{\rm fs}^q$ uses $B_q \to X_q \ell^+ \ov{\nu_\ell}$ decays,
so that $a_{\rm fs}^q$ is often called the \emph{semileptonic CP asymmetry}.

\boldmath
\section{$\dm_d$ and $\dm_s$}
\unboldmath
In order to predict $\dm_q$ we need to compute $M_{12}^q$.
Key quantities entering $M_{12}^q$ are the CKM element $V_{tq}$, the top
mass $\ov m_t$ and  $f_{B_q}^2 B$, which parameterises the matrix element 
\begin{eqnarray}
\bra{B_q} Q \ket{\ov B_q}  &=& \frac{8}{3} M^2_{B_q}\, f^2_{B_q} B 
      \label{defb} 
\end{eqnarray}
of the four-quark operator ($\alpha,\beta=1,2,3$ are colour indices): 
\begin{eqnarray}
Q & =&   \ov q_\alpha \gamma_\mu (1-\gamma_5) b_\alpha \, 
         \ov q_\beta \gamma^\mu (1-\gamma_5) b_\beta . 
\label{defq} 
\end{eqnarray} 
While the decay constants $f_{B_d}$ and $f_{B_s}$ differ numerically,
no non-perturbative computation of the bag factor $B$ has shown a
significant difference for $B_d$ and $B_s$ mesons. 
 
Updating $\dm_q$ computed in \cite{bjw} to $m_t^{\rm pole} = 171.4 \pm
2.1 \, {\rm GeV}$ \cite{topmass}, which corresponds to $\ov m_t (\ov
m_t) = 163.8 \pm 2.0\, {\rm GeV}$ in the $\ov{\rm MS}$ scheme, gives
\begin{eqnarray}
\dm_d^{\rm SM} & = & (0.53\pm 0.02) \, \mbox{ps}^{-1}
          \lt( \frac{|V_{td}|}{0.0082}\rt)^2  \nn
&&\qquad\qquad      \cdot    \lt( \frac{f_{B_d}}{200 \, \mbox{MeV}} \rt)^2
             \frac{B}{0.85} \nn
\dm_s^{\rm SM} & = & \lt( 19.3 \pm 0.6 \rt) \, \mbox{ps}^{-1}
          \lt( \frac{|V_{ts}|}{0.0405}\rt)^2  \nn 
&& \qquad\qquad  \cdot
          \lt( \frac{f_{B_s}}{240 \, \mbox{MeV}} \rt)^2
             \frac{B}{0.85} \label{dms}
\end{eqnarray}
$B$ in \eq{dms} and all other bag factors appearing in this talk are
evaluated in the $\ov{\rm MS}$ scheme at the scale $\mu=\ov m_b$. The
frequently used scheme-invariant bag paramter $\widehat B$ is larger
than $B$ by a factor of 1.5.  Both $\dm_d$ and $\dm_s$ are
well-measured. Here we concentrate on the phenomenological impact of
this year's discovery of \bbms\ and the precise measurement of $\dm_s$
\cite{ichep06dm}:
\begin{eqnarray}
17 \, \mbox{ps}^{-1} \!\! & \leq &\!\!  \dm_s \, \leq \, 21 \, \mbox{ps}^{-1}
  \quad @90\% \, \mbox{CL}  \quad \mbox{(D\O)} \nn
\dm_s \!\! & =&\!\!  17.77\pm{0.10} 
       \pm 0.07 \, \mbox{ps}^{-1} \, 
   \; \mbox{(CDF)} . \; \label{dmexp}
\end{eqnarray} 
$|V_{cb}|=0.0415 \pm 0.0010$ determines $|V_{ts}|=0.0405 \pm 0.0010 $,
so that CKM uncertainties are not an issue for $\dm_s$. However,
non-perturbative computations of $f_{B_s}$ and $B$ still cover a wide
range. The ballpark of results from lattice QCD \cite{lattice} and QCD
sum rules \cite{sum} is represented by $f_{B_s}=240 \pm 40 \,
\mbox{MeV}$ and $B = 0.85 \pm 0.06$, which implies   
\begin{eqnarray}
\dm_s^{\rm SM} & = & \lt( 19.30 \pm 6.68 \rt) \, \mbox{ps}^{-1}
\label{dmsnum}
\end{eqnarray}
\boldmath
\section{$\dg_d$ and $\dg_s$}
\unboldmath
In order to predict $\dg_q$ we need to compute $\Gamma_{12}^q$. 
There are two important differences compared to $M_{12}^q$: 
first, $\Gamma_{12}^q$ involves an operator product expansion at the
scale $m_b$, so that one faces two expansion parameters, $\ov
\Lambda/m_b$ and $\alpha_s(m_b)$, where $\ov \Lambda \sim (M_{B_q}-m_b)$
is the relevant hadronic scale of the problem. Second, even in the
leading order of $\ov \Lambda/m_b$ the prediction of 
$\Gamma_{12}^q$ involves two operators. In addition to $Q$ in \eq{defq}
one encounters
\begin{eqnarray}
Q_S & = &  \ov{q}_\alpha (1+\gamma_5)  b_\alpha \, 
           \ov{q}_\beta  (1+\gamma_5)  b_\beta ,
  \label{defqs}
\end{eqnarray}
which involves a new bag parameter $B_{S}^\prime$ in the range
$B_S^\prime \; =\; 1.34 \pm 0.12$ \cite{lattice}.  $\Gamma_{12}^q$ is
known to next-to-leading-order (NLO) in both $\lbar/m_b$ \cite{bbd1} and
$\alpha_s(m_b)$ \cite{bbgln1,rome03,bbln}. With current values of the
input parameters (most relevant are $\ov m_b( \ov m_b) = 4.22 \pm 0.08
\, {\rm GeV}$, $\ov m_c( \ov m_c) = 1.30 \pm 0.05 \, {\rm GeV}$ and
$\alpha_s (M_Z) = 0.1189 \pm 0.0010$), the result of \cite{bbgln1} is
updated to
\begin{eqnarray}
\dg_s^{\rm old}& = & 
\left( 0.070   \pm 0.042 \right) \, \mbox{ps}^{-1}
\label{dgold}
\end{eqnarray}    
$\dg_s^{\rm old}$ is pathological in several aspects: both the
$\lbar/m_b$ and $\alpha_s$ corrections are large and negative, which
limits the accuracy in \eq{dgold}. Further $\dg_s^{\rm old}$ is
dominated by the matrix element of $Q_S$, so that hadronic uncertainties
do not cancel in the ratio $\dg_s^{\rm old}/\dm_s \propto B_S^\prime/B$.

\begin{figure*}
\centerline{
\includegraphics[height=0.34\textheight,angle=0]{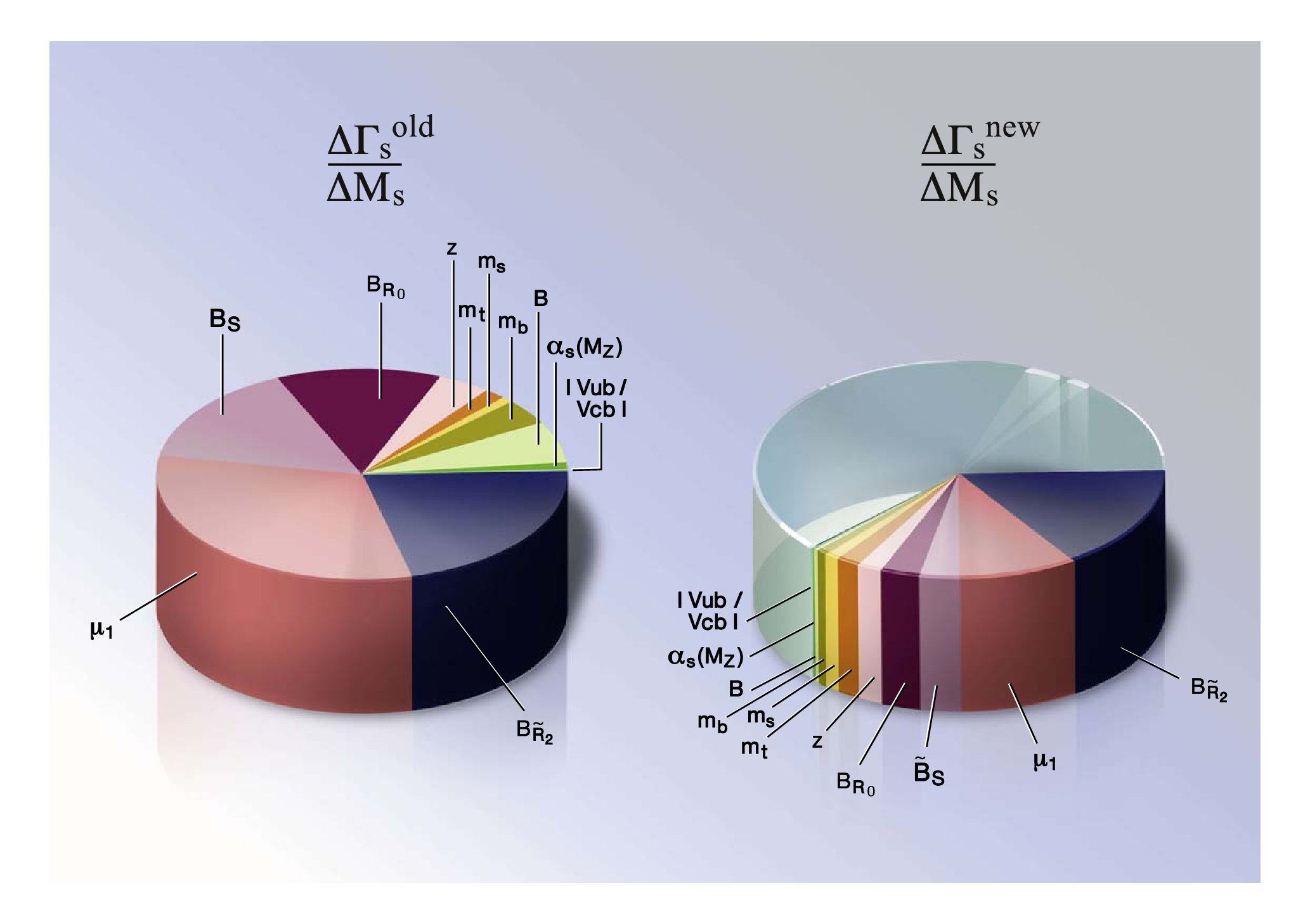}}
\caption{Uncertainty budget for $\dg_s/\dm_s$. 
  The largest uncertainties stem from the renormalisation scale $\mu_1$ of
  the $\Delta B=1$ operators and the bag parameter 
  $B_{\widetilde{R}_2}=1.0 \pm 0.5$ of
  one of the $1/m_b$--suppressed operators. 
  $B_{R_0}=1.0 \pm 0.5 $ is the bag parameter of 
  $R_0$ in \eq{defr0} and $z=\ov m_c^2/\ov m_b^2$.
  The transparent segment of
  the right pie chart shows the improvement with respect to the old
  result on the left.  }\label{Kuchendgdm}
\end{figure*}

When computing the leading contribution to $\Gamma_{12}^q$, one first 
encounters a third operator,
\begin{eqnarray}
\widetilde{Q}_S & = &  \ov{q}_\alpha (1+\gamma_5)  b_\beta \, 
                       \ov{q}_\beta (1+\gamma_5)  b_\alpha .
  \label{defqst}
\end{eqnarray}
Subsequently $\widetilde{Q}_S$ is traded for 
\begin{eqnarray}  
R_0 &\equiv&                            Q_S 
              \; + \; \alpha_1  \tilde Q_S 
           \; + \; \frac{1}{2}  \alpha_2 Q, 
\label{defr0}
\end{eqnarray}  
which belongs to the sub-leading order in $\ov \Lambda /m_b$. $\alpha_1$
and $\alpha_2$ are QCD factors \cite{bbgln1,bbln,ln}. 
Writing 
\begin{eqnarray}  
\bra{B_s} \widetilde Q_S \ket{\ov B_s} &=& \frac{1}{3}  M^2_{B_s}\,
                  f^2_{B_s} \widetilde B_S^\prime , 
    \label{defbstp} 
\end{eqnarray}
a lattice computation finds $\widetilde B_S^\prime \; =\; 1.41 \pm 0.12$
\cite{bgmpr}. With this result we find the matrix element of
$\widetilde{Q}_S$ roughly five times smaller than that of $Q_S$, which
involves an overall factor of --5/3 instead of 1/3 in \eq{defbstp}. We
propose to use \eq{defr0} to eliminate $Q_S$ from the operator basis, so
that the leading term of the operator product expansion involves $Q$ and
$\widetilde Q_S$. This change of basis reshuffles the expansions in both
$\lbar/m_b$ and $\alpha_s(m_b)$. Since terms of order $\alpha_s(m_b)
\cdot \lbar/m_b$ have not been calculated yet, the central value of our
prediction for $\dg_s$ changes within the error bar of the previous
predictions in \cite{bbgln1,rome03}. 
In our new basis we find that all pathologies    
vanish: the $\lbar/m_b$ and $\alpha_s(m_b)$ corrections
have shrunk to their natural sizes and $\dg_s$ is now dominated 
by the matrix element of $Q$:
\begin{eqnarray}
\dg_s^{\rm SM} & = & \left( \frac{f_{B_s}}{240 \, \mbox{MeV}} \right)^2
\big[ \, (0.105 \pm 0.016) B 
\nn
&& \!\!\!\!+\,  (0.024 \pm 0.004)  \widetilde B_S^\prime 
 \, -\, 0.027 \big] \, \mbox{ps}^{-1}   \;
\label{finaldg}
\end{eqnarray}
The ratio $\dg_s/\dm_s$ is now almost free from hadronic uncertainties.
Inserting the numerical ranges for $B$ and $\widetilde B_S^\prime $
quoted before \eq{dmsnum} and after \eq{defbstp} yields:
\begin{eqnarray}
\frac{\dg_s^{\rm SM}}{\dm_s^{\rm SM}}  & = & 
\left( 49.7 \pm 9.4 \right) \cdot 10^{-4} . \label{dgdmnum}
\end{eqnarray}
The progress due to the change of the operator basis is depicted in 
\fig{Kuchendgdm}.
Using the CDF measurement of $\dm_s$ in \eq{dmexp} we predict from
\eq{dgdmnum}:
\begin{eqnarray}
\dg_s^{\rm SM} & = & 
0.088 \pm 0.017\, \mbox{ps}^{-1}   \nn
\frac{\dg_s^{\rm SM}}{\Gamma_s} & =&   0.127 \pm 0.024 \, . \label{dgsnum}
\end{eqnarray}
Here $\Gamma_s$ is the average width of the two $B_s$ mass eigenstates
and the theoretical relation $\Gamma_s=1/\tau_{B_d} (1+{\cal O}(0.01))$
has been used. Any experimental violation of \eq{dgsnum} will signal new
physics in either $\dm_s$ or $\dg_s$. The corresponding results for the 
$B_d$ system are
\begin{eqnarray}
\frac{\dg_d^{\rm SM}}{\dm_d^{\rm SM}} &=& \lt( 52.6 \epm{11.5}{12.8} \rt) 
             \, \cdot 10^{-4} \nn
\dg_d^{\rm SM} &=& 
     \lt( 26.7 \epm {5.8}{6.5}\rt) \cdot 10^{-4} \; 
       {\rm ps}^{-1} \nn
\frac{\dg_d^{\rm SM}}{\Gamma_d^{\rm SM}} 
&=& \lt( 40.9 \epm{8.9}{9.9} \rt) \cdot 10^{-4},
   \label{dgdnum}
\end{eqnarray}
where $\dm_d^{\rm exp}= 0.507 \pm 0.004 \, {\rm ps}^{-1}$ and 
$\tau_{B_d}^{\rm exp}  = 1.530 \pm 0.009 $ has been used. 

\boldmath
\section{$a_{\rm fs}^d$, $a_{\rm fs}^s$, $\phi_s$ and $\phi_d$}
\unboldmath The CP asymmetries in flavour-specific decays are typically
measured using semileptonic decays. No tagging is required, $a_{\rm
  fs}^q$ can be found by simply counting positively and negatively
charged leptons. It may be worthwile to study the time evolution of
the untagged sample \cite{n},
\begin{eqnarray}
\lefteqn{
\frac{ \Gamma[ \BqorBqbar \to X^- \ell^+\nu_\ell ,t ] \, -\, 
       \Gamma[ \BqorBqbar \to X^+ \ell^- \ov \nu_\ell ,t ]}{
       \Gamma[ \BqorBqbar \to X^- \ell^+\nu_\ell ,t ] \, +\,
       \Gamma[ \BqorBqbar \to X^+ \ell^- \ov \nu_\ell ,t ]} 
\; =}\nn
&& \qquad\qquad\frac{a_{\rm fs}^q}{2} \lt[ 1- 
    \frac{\cos(\dm_q \, t)}{\cosh{(\dg_q\, t/2)}}  \rt],
\label{afqt}
\end{eqnarray}
as it may help to reject fake effects from detection asymmetries and to
separate the $B_d$ and $B_s$ samples in hadron collider experiments
through $\dm_d\neq\dm_s$.  Further, the time evolution of any (tagged or
untagged) decay of $B_q$ mesons contains information on $a_{\rm fs}^q$
\cite{n}, which might be useful to add statistics in the determination
of $a_{\rm fs}^d$ at B factories. 

Our new operator basis does not improve the Standard Model prediction of
$a_{\rm fs}^q$ over the one in \cite{bbln,rome03}. With up-to-date input
parameters we find
\begin{eqnarray}
a_{\rm fs}^{\rm SM,s} & = & \left( 2.06 \pm 0.57 \right) \cdot 10^{-5}, \nn
a_{\rm fs}^{\rm SM,d} & = & \lt(  -4.8\epm{1.0}{1.2} \rt) \, 
                \cdot 10^{-4}. \label{afsnum} 
\end{eqnarray}
$a_{\rm fs}^d$ depends on CKM parameters. The quoted value uses 
$\beta=23^\circ\pm 2^\circ$ for the angle of the unitarity triangle
measured in $B_d \to J/\psi K_S$ and $R_t =0.86\pm 0.11$ for the side 
of the triangle adjacent to $\beta$ and opposite to $\gamma$. The value
of $R_t$ ignores any input from $\dm_d/\dm_s$, since our focus is on
potential new physics in \bbms. Our predictions in \eq{afsnum}
correspond to the following values of the CP phases appearing in
\eq{defafs}:
\begin{eqnarray}
\phi_d^{\rm SM} &=&  -0.091\epm{0.026}{0.038} \; =\; 
            -5.2^\circ\epm{1.5^\circ}{2.1^\circ} \nn 
\phi_s^{\rm SM} & = & (4.2\pm 1.4 )\cdot 10^{-3}
      \; = \; 0.24^\circ \pm 0.08^\circ \label{phinum}
\end{eqnarray}

\section{$\dg_s$, $\dg_s/\dm_s $ and $a_{\rm fs}^s$ beyond
  the SM}%
\begin{figure*}
\centerline{
\includegraphics[height=0.3\textheight,angle=0]{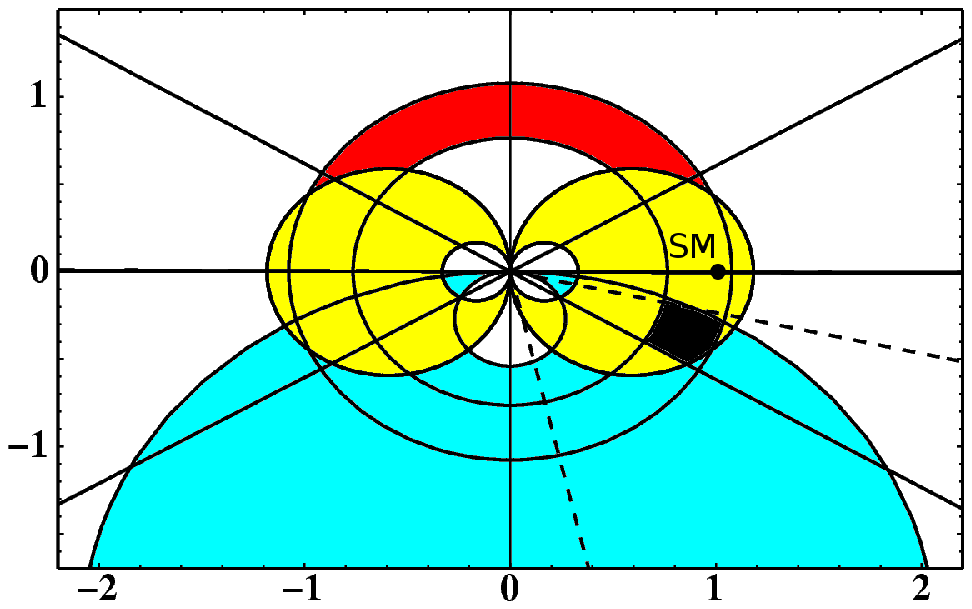}}
\caption{Current experimental bounds in the complex $\Delta_s$-plane.
  The bound from $\Delta M_s$ is the red (dark-grey) annulus around the
  origin. The bound from $|\Delta \Gamma_s|/ \Delta M_s$ corresponds to
  the yellow (light-grey) region and the bound from $a_{\rm fs}^s$ is
  given by the light-blue (grey) region. The angle $\phi_s^\Delta$ can
  be extracted from $|\Delta \Gamma_s|$ (solid lines) with a four--fold
  ambiguity --- each of the four regions is bounded by a solid ray and
  the x-axis --- or from the angular analysis in $B_s \to J / \Psi \phi$
  (dashed line). (No mirror solutions from discrete ambiguities are
  shown for the latter.)  The current experimental situation shows a
  $2\sigma$ deviation from the Standard Model case $\Delta_s=
  1$.}\label{boundbandreal}
\end{figure*}

Looking beyond the Standard Model we now allow for a new CP phase
$\phi_s^\Delta$ which adds to the Standard Model value in \eq{phinum}.
Then $a_{\rm fs}^s$ not only depends on $\imag \Gamma_{12}^s/M_{12}^s$,
but also on $\real \Gamma_{12}^s/M_{12}^s$, which we can predict in a
much better way with our new operator basis.  We parameterise the effect
of new physics (similarly to \cite{nir}) by
\begin{eqnarray}
M_{12}^s & \equiv & M_{12}^{\rm SM,s} \cdot  \Delta_s \, ,
\qquad  \Delta_s \; \equiv \;  |\Delta_s| e^{i \phi^\Delta_s} . 
 \label{defdel}
\end{eqnarray}
Then one easily finds
\begin{eqnarray}
\dm_s  & = & \dm_s^{\rm SM} \,  |\Delta_s|  \nn
&=&
(19.30 \pm 6.74 ) \, \mbox{ps}^{-1} \cdot | \Delta_s| \, ,
\nn
\Delta \Gamma_s  & = & 2 |\Gamma_{12}^s|
     \, \cos \left( \phi_s^{\rm SM} + \phi^\Delta_s \right)\nn
&=& (0.096 \pm 0.039) \, \mbox{ps}^{-1} 
\cdot \cos \left( \phi_s^{\rm SM} + \phi^\Delta_s \right)
\nn
\frac{\Delta \Gamma_s}{\Delta M_s} 
&= &
 \frac{|\Gamma_{12}^s|}{|M_{12}^{\rm SM,s}|} 
\cdot \frac{\cos \left( \phi_s^{\rm SM} + \phi^\Delta_s \right)}{|\Delta_s|}
\nn
&=&
\left( 4.97 \pm 0.94 \right) \cdot 10^{-3} 
\cdot \frac{\cos \left( \phi_s^{\rm SM} + \phi^\Delta_s \right)}{|\Delta_s|}
\nn
a_{\rm fs}^s 
&= &
 \frac{|\Gamma_{12}^s|}{|M_{12}^{\rm SM,s}|} 
\cdot 
\frac{\sin \left( \phi_s^{\rm SM} + \phi^\Delta_s \right)}{|\Delta_s|}
\label{boundafs} \\
&=& \left( 4.97 \pm 0.94 \right) \cdot 10^{-3} 
\cdot \frac{\sin \left( \phi_s^{\rm SM} + \phi^\Delta_s
  \right)}{|\Delta_s|} \, .
\no
\end{eqnarray}
A further source of information on the phase $\phi_s^{\Delta}$ is the
angular distribution of the decay $\BsorBsbar \to J/\psi \phi$, which
contains a CP-odd term \cite{ddlr,dfn}. Recently, the D\O\ collaboration
found $ \phi_s^{\Delta}-2\beta_s = -0.79 \pm 0.56 \pm 0.01$ (and mirror
solutions in other quadrants of the complex $\Delta_s$ plane) from this
analysis \cite{dgexpnew}. Here $\beta_s= 0.020 \pm 0.005 = 1.1^\circ \pm
0.3^\circ$ is the relevant combination of CKM phases.  Combining all
available experimental information with our new theory predictions (see
\cite{ln} for details) gives the constraints depicted in
\fig{boundbandreal}. Setting $|\Delta_s|$ to its Standard Model value
$|\Delta_s|=1$ we find with the data from \cite{dgexpnew} and
\cite{dzexp}:
\begin{eqnarray}
\sin (\phi_s^{\Delta}-2\beta_s)  &=& 
-0.77 \pm 0.04{}_{\mbox{\scriptsize (th)}} 
                \pm 0.34 {}_{\mbox{\scriptsize (exp)}}, \no
\end{eqnarray}
which deviates from the Standard Model value
$\sin(-2\beta_s)=-0.04\pm0.01$ by 2 standard deviations. 

\section{Conclusions}
We have improved the theoretical prediction of $\real
\Gamma_{12}^q/M_{12}^q$ (with $q=d$ or $q=s$) by introducing a new
operator basis. This quantity enters the predictions of the width
difference $\dg_q$ in the \bbq\ system and the prediction of 
the CP asymmetry in flavour-specific decays, $a_{\rm fs}^q$, in scenarios 
of physics beyond the Standard Model. Applying our formulae to D\O\ data
we find that the \bbms\ phase deviates from the Standard Model value 
by $2\sigma$. While this is not statistically significant, it shows 
that current experiments are reaching the sensitivity to probe any new
physics entering the \bbms\ phase.



\begin{thebibliography}{99}
\bibitem{ckm} N.~Cabibbo,
Phys.\ Rev.\ Lett.\  {\bf 10}, 531 (1963);
M.~Kobayashi and T.~Maskawa,
Prog.\ Theor.\ Phys.\  {\bf 49}, 652 (1973).

\bibitem{ichep06} The experimental situation is summarised in:
  M.~Hazumi, plenary talk at \emph{33rd International Conference On High
    Energy Physics (ICHEP 06)}, 26 Jul -- 2 Aug 2006, Moscow, Russia. 

\bibitem{cmm}
  D.~Chang, A.~Masiero and H.~Murayama,
  Phys.\ Rev.\ D {\bf 67}, 075013 (2003)
  [arXiv:hep-ph/0205111].

\bibitem{jn} Effects of the model in \cite{cmm} on \bbs\ mixing have
  been studied in:
   S.~J\"ager and U.~Nierste,
  Eur.\ Phys.\ J.\ C {\bf 33}, S256 (2004)
  [arXiv:hep-ph/0312145];
   S.~J\"ager and U.~Nierste,
  in \emph {Proceedings of the 12th International Conference On
  Supersymmetry And Unification Of Fundamental Interactions (SUSY 04)}, 
  17-23 Jun 2004, Tsukuba, Japan, p.\ 675-678, 
  Ed.~K. Hagiwara, J. Kanzaki, N. Okada 
  [hep-ph/0410360];
    S.~J\"ager,
  hep-ph/0505243, to appear in the 
\emph{Proceedings of the XLth Rencontres
    de Moriond, Electroweak Interactions and Unified Theories}, 5-12 Mar
  2005, La Thuile, Aosta Valley, Italy, Ed. J. Tran Thanh Van.

\bibitem{ln}
  A.~Lenz and U.~Nierste,
  arXiv:hep-ph/0612167.


\bibitem{run2}
K.~Anikeev {\it et al.},
\emph{$B$ physics at the Tevatron: Run II and beyond},
[hep-ph/0201071], Chapters 1.3 and 8.3.



\bibitem{bjw}
A.J. Buras, M. Jamin and P.H. Weisz, Nucl. Phys. {\bf B347}, 491 (1990).


\bibitem{topmass}
  E.~Brubaker {\it et al.}  [Tevatron Electroweak Working Group],
  arXiv:hep-ex/0608032.
  
\bibitem{ichep06dm} 
  Talks by D.~Glenzinski (plenary), T.~Moulik and S.~Giagu at \emph{33rd
    International Conference On High Energy Physics (ICHEP 06)}, 26 Jul
  -- 2 Aug 2006, Moscow, Russia.  Talks by P.  Tamburello [D\O] and
  A.~Belloni [CDF] at \emph{Beauty 2006}, 25-29 Sep 2006, Oxford,
  England, Nucl.\ Phys.\ Proc.\ Suppl.\ {\bf 170} (2007) 123 and
  \emph{ibid.}\ 129.

\bibitem{lattice}
S.~Hashimoto and T.~Onogi,
arXiv:hep-ph/0407221;
S.~Hashimoto,
Int.\ J.\ Mod.\ Phys.\ A {\bf 20} (2005) 5133
[arXiv:hep-ph/0411126];
M.~Okamoto,
PoS {\bf LAT2005} (2006) 013
[arXiv:hep-lat/0510113].
A.~Ali Khan {\it et al.}  [CP-PACS Collaboration],
Phys.\ Rev.\ D {\bf 64} (2001) 054504
[arXiv:hep-lat/0103020];
A.~Ali Khan {\it et al.}  [CP-PACS Collaboration],
Phys.\ Rev.\ D {\bf 64} (2001) 034505 [arXiv:hep-lat/0010009];
C.~Bernard {\it et al.}  [MILC Collaboration],
Phys.\ Rev.\ D {\bf 66} (2002) 094501 [arXiv:hep-lat/0206016];
S.~Collins, C.~T.~H.~Davies, U.~M.~Heller, A.~Ali Khan, J.~Shigemitsu, 
J.~H.~Sloan and C.~Morningstar,
Phys.\ Rev.\ D {\bf 60} (1999) 074504 [arXiv:hep-lat/9901001].
S.~Aoki {\it et al.}  [JLQCD Collaboration],
%
Phys.\ Rev.\ Lett.\ {\bf 91} (2003) 212001 [arXiv:hep-ph/0307039].
V.~Gimenez and J.~Reyes,
Nucl.\ Phys.\ Proc.\ Suppl.\ {\bf 94} (2001) 350
[arXiv:hep-lat/0010048].  N.~Yamada {\it et al.}  [JLQCD Collaboration],
%
Nucl.\ Phys.\ Proc.\ Suppl.\ {\bf 106} (2002) 397
[arXiv:hep-lat/0110087].
M.~Wingate, C.~T.~H.~Davies, A.~Gray, G.~P.~Lepage and J.~Shigemitsu,
Phys.\ Rev.\ Lett.\ {\bf 92} (2004) 162001 [arXiv:hep-ph/0311130];
A.~Gray {\it et al.}  [HPQCD Collaboration],
Phys.\ Rev.\ Lett.\ {\bf 95} (2005) 212001 [arXiv:hep-lat/0507015].
  E.~Dalgic {\it et al.},
  arXiv:hep-lat/0610104;
 J. Shigemitsu for HPQCD Collaboration, talk at LATTICE 2006,
  http://www.physics.utah.edu/lat06/ 
  abstracts/sessions/weak/s1//Shige\-mit\-su\-$\_$Jun\-ko.pdf.


\bibitem{sum}
M.~Jamin and B.~O.~Lange,
Phys.\ Rev.\ D {\bf 65} (2002) 056005
[arXiv:hep-ph/0108135].
J.~G.~Korner, A.~I.~Onishchenko, A.~A.~Petrov and A.~A.~Pivovarov,
%
Phys.\ Rev.\ Lett.\  {\bf 91} (2003) 192002
[arXiv:hep-ph/0306032].

\bibitem{bbd1}
M. Beneke, G. Buchalla and I. Dunietz, 
Phys. Rev. {\bf D54}, 4419 (1996).

\bibitem{bbgln1}
M.~Beneke, G.~Buchalla, C.~Greub, A.~Lenz and U.~Nierste,
Phys.\ Lett.\ B {\bf 459} (1999) 631
[arXiv:hep-ph/9808385].

\bibitem{rome03}
M.~Ciuchini, E.~Franco, V.~Lubicz, F.~Mescia and C.~Tarantino,
JHEP {\bf 0308} (2003) 031
[arXiv:hep-ph/0308029].

\bibitem{bbln}
M.~Beneke, G.~Buchalla, A.~Lenz and U.~Nierste,
Phys.\ Lett.\ B {\bf 576} (2003) 173
[arXiv:hep-ph/0307344].

\bibitem{bgmpr}
D.~Becirevic, V.~Gimenez, G.~Martinelli, M.~Papinutto and J.~Reyes,
JHEP {\bf 0204} (2002) 025
[arXiv:hep-lat/0110091].

\bibitem{n} 
  U.~Nierste, 
  [hep-ph/0406300], in: \emph{Proceedings of the XXXIXth Rencontres
    de Moriond, Electroweak Interactions and Unified Theories}, 21-28
  Mar 2004, La Thuile, Aosta Valley, Italy, Ed. J. Tran Thanh Van.

\bibitem{nir}
  Y.~Grossman, Y.~Nir and M.~P.~Worah,
  Phys.\ Lett.\ B {\bf 407} (1997) 307
  [arXiv:hep-ph/9704287].

\bibitem{ddlr}
  A.~S.~Dighe, I.~Dunietz, H.~J.~Lipkin and J.~L.~Rosner,
  Phys.\ Lett.\ B {\bf 369} (1996) 144
  [arXiv:hep-ph/9511363].
    A.~S.~Dighe, I.~Dunietz and R.~Fleischer,
  Eur.\ Phys.\ J.\ C {\bf 6} (1999) 647
  [arXiv:hep-ph/9804253].

\bibitem{dfn}
  I.~Dunietz, R.~Fleischer and U.~Nierste,
  Phys.\ Rev.\ D {\bf 63} (2001) 114015
  [arXiv:hep-ph/0012219].

\bibitem{dgexpnew}
D\O\ collaboration, conference note 5144, http://www-do.fnal.gov/.


\bibitem{dzexp}
D\O\ collaboration, conference note 5143, http://www-do.fnal.gov/.
V.~Abazov  [D\O\ Collaboration], arXiv:hep-ex/0609014.


\end{thebibliography}
\end{document}